\documentclass[prd,twocolumn,eqsecnum,showpacs,amsfonts,amssymb]{revtex4}

\usepackage{graphicx}

\usepackage{bm}

\newcommand{\beq}{\begin{equation}}
\newcommand{\eeq}{\end{equation}}
\newcommand{\beqs}{\begin{eqnarray}}
\newcommand{\eeqs}{\end{eqnarray}}

\begin{document}

\title{Study of Scheme Transformations to Remove Higher-Loop
  Terms in the $\beta$ Function of a Gauge Theory} 

\author{Robert Shrock}

\affiliation{C. N. Yang Institute for Theoretical Physics \\
Stony Brook University \\
Stony Brook, NY 11794 }

\begin{abstract}

Since three-loop and higher-loop terms in the $\beta$ function of a gauge
theory are scheme-dependent, one can, at least for sufficiently small coupling,
carry out a scheme transformation that removes these terms.  A basic question
concerns the extent to which this can be done at an infrared fixed point of an
asymptotically free gauge theory. This is important for quantitative analyses
of the scheme dependence of such a fixed point.  Here we study a scheme
transformation $S_{R,m}$ with $m \ge 2$ that is constructed so as to remove the
terms in the beta function at loop order $\ell=3$ to $\ell=m+1$, inclusive.
Starting from an arbitrary initial scheme, we present general expressions for
the coefficients of terms of loop order $\ell$ in the beta function in the
transformed scheme from $\ell=m+2$ up to $\ell=8$.  Extending a previous study
of $S_{R,2}$, we investigate the range of applicability of the $S_{R,3}$ scheme
transformation in an asymptotically free SU($N_c$) gauge theory with an
infrared zero in $\beta$ depending on the number, $N_f$, of fermions in the
theory. We show that this $S_{R,3}$ scheme transformation can only be applied
self-consistently in a restricted range of $N_f$ with a correspondingly small
value of infrared fixed-point coupling.  We also study the effect of
higher-loop terms on the beta function of a U(1) gauge theory.

\end{abstract}


\maketitle

% =======================================================================

\section{Introduction}
\label{intro}

The evolution of the coupling $g(\mu)$ as a function of the reference Euclidean
momentum scale, $\mu$, from the ultraviolet (UV) to the infrared (IR) in an
asymptotically free gauge theory is of fundamental field-theoretic importance.
This evolution of $g(\mu)$, or equivalently, $\alpha(\mu) = g(\mu)^2/(4\pi)$,
is described by the $\beta$ function of the theory.  Terms at loop order $\ell
\ge 3$ in the $\beta$ function are dependent on the scheme used for
regularization and renormalization. Hence, one expects that, at least for
sufficiently small coupling, it is possible to carry out a scheme
transformation that removes these terms and yields a $\beta$ function with only
one- and two-loop terms \cite{thooft77}. In \cite{sch} we constructed what is,
to our knowledge, the first explicit scheme transformation that removes terms
at loop order $\ell \ge 3$ from the beta function, at least in the vicinity of
the UV fixed point at $\alpha=0$. 

An important application of such a scheme transformation is to asymptotically
free gauge theories that have an infrared zero in the $\beta$ function.
Depending on how large the value of the coupling is at this IR zero, it is
either an exact or approximate fixed point of the renormalization group of the
theory.  In order to understand the physical implications of this IR zero, it
is necessary to assess the effect of scheme dependence on its value. Hence, a
crucial question concerning a scheme transformation designed to remove terms at
three- and higher-loop order in the beta function, is whether one can use it in
the vicinity of an IR zero of this function. Indeed, a scheme transformation
that is acceptable for small coupling can produce unphysical effects that
render it inapplicable for somewhat larger couplings \cite{sch}.

Here we study a scheme transformation $S_{R,m}$ with $m \ge 2$ that is
constructed so as to remove the $\ell$-loop terms in the beta function at loop
order $\ell=3$ to $\ell=m+1$, inclusive. We investigate this to the
highest-loop order possible using known coefficients of $\beta$ for a
non-Abelian gauge theory, namely $\ell=4$ loop order, corresponding to
$m=3$. We focus on an asymptotically free gauge theory with gauge group
SU($N_c$) containing $N_f$ massless fermions in the fundamental representation,
although many of our results apply to the case of an arbitrary gauge group $G$
with $N_f$ massless fermions in a general representation $R$ of $G$ \cite{mf}.
Starting from an arbitrary initial scheme, we present general expressions for
the coefficients of terms of loop order $\ell$ in the beta function in the
transformed scheme from $\ell=m+2$ up to $\ell=8$.  It was shown in \cite{sch}
that the $S_{R,2}$ scheme transformation has a limited range of applicability
and cannot be used for a substantial subset of $N_f$ values where the theory
has an IR zero in $\beta$ because it produces unphysical effects, namely a
reversal of the sign of $\alpha$.  This finding naturally leads to a question:
how general is this problem and can one alleviate or circumvent it by using the
higher-order scheme transformation $S_{R,3}$?

We address and answer this question here. We will show here that the problem is
generically still present with the $S_{R,3}$ scheme transformation.  For
example, we will show that for a theory with gauge group SU(3) and $N_f=12$
fermions, one cannot use the $S_{R,3}$ scheme transformation in the vicinity of
the (scheme-independent) IR zero of the two-loop $\beta$ function because it
produces the same type of unphysical results that the $S_{R,2}$ transformation
does. Thus, while it is true that one can remove terms at loop order $\ell \ge
3$ in a $\beta$ function for sufficiently small $\alpha$, one must take
considerable care in attempting such a scheme transformation at moderate values
of $\alpha$ relevant for a generic infrared fixed point. As part of our work,
we also discuss some higher-loop properties of the $\beta$ function and
associated issues of scheme dependence for a U(1) gauge theory.

This paper is organized as follows. In Sect. \ref{background} we give some
additional background and motivations for the current work.  The definition and
some properties of a general scheme transformation are presented in
Sect. \ref{st}. In Sect. \ref{srm} we define the scheme transformation
$S_{R,m}$ that, at least for sufficiently small $\alpha$, removes terms in the
beta function from loop order $\ell=3$ to $\ell=m+1$, inclusive.  Explicit
expressions for the resultant coefficients in the new scheme are presented in
Sect. \ref{bpsrm}. In Sect. \ref{sr2_application} we discuss the range of
applicability of the transformation $S_{R,2}$ at an IR zero of the beta
function.  In Sect. \ref{sr3_application} we present our results on the range
of applicability of the $S_{R,3}$ scheme transformation. As discussed in
Sect. \ref{lnn_section}, further insights concerning the range of applicability
of these scheme transformations are gained by studying the limit of an
SU($N_c$) gauge theory with $N_f$ fermions in the fundamental representation in
the limit $N_c \to \infty$, $N_f \to \infty$ with the ratio $N_f/N_c$ fixed.
In Sect. \ref{u1} we discuss some higher-loop properties of the $\beta$
function for a U(1) gauge theory.  Our conclusions are given in
Sect. \ref{conclusions}, and some additional relevant formulas are listed in
several appendices.

% =========================================================================

\section{Background}
\label{background} 

The dependence of $\alpha(\mu)$ on $\mu$ is described by the $\beta$ function
\cite{rg,dgdt}
\beq
\beta \equiv \beta_\alpha = \frac{d\alpha}{dt} \ . 
\label{betadef}
\eeq
It will be convenient to introduce the quantity $a(\mu) \equiv
\alpha(\mu)/(4\pi) = g(\mu)^2/(16\pi^2)$ (the argument $\mu$ will often be
suppressed in the notation).  The $\beta$ function has the expansion
\beq
\beta_\alpha = -2\alpha \sum_{\ell=1}^\infty b_\ell \, a^\ell =
-2\alpha \sum_{\ell=1}^\infty \bar b_\ell \, \alpha^\ell \ ,
\label{beta}
\eeq
where
\beq
\bar b_\ell = \frac{b_\ell}{(4\pi)^\ell} \ .
\label{bellbar}
\eeq
The $n$-loop $\beta$ function is given by Eq. (\ref{beta}) with the upper limit
on the $\ell$ loop summation equal to $n$ instead of $\infty$. The one-loop and
two-loop coefficients, $b_1$ and $b_2$, are independent of the scheme used for
regularization and renormalization, while $b_\ell$ with $\ell \ge 3$ are
scheme-dependent \cite{gross75}. The coefficients $b_1$ and $b_2$ were
calculated for a non-Abelian Yang-Mills gauge theory in \cite{b1} and
\cite{b2,jones}.  Dimensional regularization \cite{dimreg} and minimal
subtraction \cite{ms} are particularly convenient for these loop calculations.
Calculations of $b_3$ and $b_4$ in the modified minimal subtraction scheme,
denoted $\overline{\rm MS}$ \cite{msbar}, were given in \cite{b3,b4,math}. We
recall that perturbative expansions in quantum field theory, such as
Eq. (\ref{beta}), are, in general, asymptotic expansions rather than Taylor
series expansions with finite radii of convergence.  However, a wealth of
experience with the use of perturbation theory for calculations of electroweak
cross sections and decay rates and for perturbative calculations in quantum
chromodynamics (QCD), has shown that these expansions can give reasonably
accurate results and that this accuracy increases when one carries these
computations to higher-loop order.  Extensive studies have been performed on
the scheme-dependence and related scale-dependence of perturbative QCD
calculations \cite{brodskyschemes}. 

If an asymptotically free gauge theory has sufficiently many massless fermions,
the $\beta$ function can exhibit an IR zero at a certain value, denoted
$\alpha_{IR}$ \cite{b2,bz,btd}.  If $\alpha_{IR}$ is sufficiently small, then
this is an exact IR fixed point (IRFP) of the renormalization group, and the UV
to IR evolution can be computed with reasonable accuracy, since the theory
starts with weak coupling in the deep UV and never becomes strongly coupled.
As the number of fermions, $N_f$, is decreased, $\alpha_{IR}$ increases.  For a
theory with sufficiently few fermions, as $\mu$ decreases past a scale denoted
as $\Lambda$, $\alpha(\mu)$ becomes large enough to trigger the formation of
bilinear fermion condensates that break the global chiral symmetry.  In a
vectorial gauge theory, these condensates are gauge-invariant, while in a
chiral gauge theory, if the condensates form, then they generically break the
gauge symmetry \cite{chiralgt}. Henceforth, for simplicity, we focus on the
case of a vectorial gauge theory. Associated with this condensate formation,
the fermions involved in the condensates gain dynamical masses of order
$\Lambda$. In the low-energy effective field theory applicable at scales $\mu <
\Lambda$, one integrates out these now-massive fermions, and the $\beta$
function reverts to that of a pure gauge theory, which has no (perturbative) IR
zero. Thus, in this case the formal IR zero in $\beta$ is only approximate.  As
$N_f$ decreases through a critical number, $N_{f,cr}$, the theory can be
regarded as undergoing a (zero-temperature) phase transition from chirally
symmetric to chirally broken infrared behavior \cite{chipt}. If $\alpha_{IR}$
is only slightly greater than the critical value for fermion condensation, then
the theory exhibits a slowly running coupling and associated quasi-scale
invariant behavior.

  To investigate the properties of a theory with an IR fixed point at moderate
coupling, it is necessary to calculate the value of $\alpha_{IR}$ to
higher-loop order \cite{gkgg}.  This was done up to four-loop order for
$\alpha_{IR}$ and the anomalous dimension, $\gamma_m$, of the fermion bilinear
for a general gauge group and fermion representation in \cite{bvh,ps}.  Further
higher-loop results on structural properties of $\beta$ were calculated in
\cite{bfs}-\cite{lnn}. Because the coefficients $b_\ell$ for $\ell \ge 3$ are
scheme-dependent, it is necessary to assess quantitatively how important the
effect of this scheme dependence is on the location of $\alpha_{IR}$.  This
task was carried out in \cite{sch}.  To do this, one constructs a scheme
transformation, applies it, calculates the value of $\alpha'_{IR}$ in the new
scheme, and determines how much $\alpha'_{IR}$ differs from $\alpha_{IR}$ to a
given loop order.

However, one encounters a significant complication in this program of
constructing and performing various scheme transformations at an IR zero of
$\beta$ and determining how much they shift the location of the zero.  As was
pointed out in \cite{sch}, in general, a scheme transformation that is
acceptable in the vicinity of the ultraviolet (UV) fixed point at $\alpha=0$
can produce unphysical effects in the vicinity of an infrared fixed
point. These include, for example, having an inverse that maps a (real,
positive) coupling to a negative or complex value. A set of conditions that a
scheme transformation must satisfy in order to be physically acceptable was
given \cite{sch} and was shown to be rather restrictive at a generic IR zero of
$\beta$.  A simple example is provided by the one-parameter family of scheme
transformations
\beq
a = \frac{\tanh (ra')}{r} 
\label{sth}
\eeq
(dependent on a parameter $r$), with inverse
\beq
a' = \frac{1}{2r} \, \ln \bigg ( \frac{1+ra}{1-ra} \bigg ) \ . 
\label{tanh_inverse}
\eeq
For example, if $r=4\pi$, then Eq. (\ref{sth}) is the scheme transformation
$\alpha=\tanh \alpha'$, and Eq. (\ref{tanh_inverse}) is its inverse,
$\alpha'=(1/2)\ln[(1+\alpha)/(1-\alpha)]$.  The scheme transformation
(\ref{sth}) is acceptable for small $\alpha$ and hence $a$, but if $a > 1/r$ 
(i.e., $\alpha > 4\pi/r$), then the transformation (\ref{tanh_inverse}) maps 
a physical $\alpha$ to a complex, unphysical $\alpha'$, and hence is
unacceptable. 

In addition to the general field-theoretic interest in understanding the
evolution of a gauge coupling as a function of Euclidean momentum scale, one of
the motivations for understanding the effect of scheme transformations on the
beta function that describes this is to provide further information from
continuum calculations to combine with information obtained from lattice
computations. Indeed, an intensive program of research is underway using
simulations of lattice gauge theories to study the infrared properties of gauge
theories with multiple fermions in various representations of the gauge group
\cite{lgtconf}.  In this context, it has been useful to compare results from
higher-loop continuum calculations with lattice measurements, e.g. on the
anomalous dimension of the fermion bilinear operator, $\gamma_m(\alpha)$
evaluated at $\alpha_{IR}$, making use of of the higher-loop calculations of
this IR zero of $\beta$ in \cite{bvh,ps,bc}. In the chirally symmetric IR
phase, a hypothetical all-orders calculation of $\gamma_m$ evaluated at an
all-orders calculation of $\alpha_{IR}$ would be an exact property of the
theory, while in the phase with spontaneous chiral symmetry breaking, just as
the IR zero of $\beta$ is only an approximate IR fixed point, so also,
$\gamma_m$ is only approximate, describing the running of $\bar\psi\psi$ and
the dynamically generated fermion mass near the zero of $\beta$.  In both the
chirally symmetric and chirally broken phases, one necessarily encounters the
issue of scheme dependence in the calculation of both $\alpha_{IR}$ and
$\gamma_m$ evaluated at $\alpha=\alpha_{IR}$ at a finite loop order $\ell \ge
3$.  It is therefore necessary to understand as well as possible the effects of
scheme transformations, in particular, the scheme transformation $S_{R,m}$ that
can remove terms in the $\beta$ function from loop order $\ell=3$ to
$\ell=m+1$.  We proceed to discuss these scheme transformations next.

% ======================================================================

\section{General Framework for Scheme Transformations} 
\label{st}

A scheme transformation can be expressed as a mapping between $\alpha$ and
$\alpha'$, or equivalently, $a$ and $a'$, which we write as $a = a' f(a')$.
We will refer to $f(a')$ as the scheme transformation function.  To keep the UV
properties the same, one requires that $f(0) = 1$.  We will consider $f(a')$ 
that are analytic about $a=a'=0$ and hence can be expanded in the form
\beq
f(a') = 1 + \sum_{s=1}^{s_{max}} k_s (a')^s =
        1 + \sum_{s=1}^{s_{max}} \bar k_s (\alpha')^s \ ,
\label{faprime}
\eeq
where the $k_s$ are constants, $\bar k_s = k_s/(4\pi)^s$, and, {\it a priori},
$s_{max}$ may be finite or infinite.  From Eq. (\ref{faprime}), it
follows that the Jacobian $J=da/da' = d\alpha/d\alpha'$ satisfies
$J=1$ at $a=a'=0$. After the scheme transformation is applied, the beta 
function in the new scheme has the form (\ref{beta}) with a new set of
coefficients, $b'_\ell$, 
\beq
\beta_{\alpha'} \equiv \frac{d\alpha'}{dt} = \frac{d\alpha'}{d\alpha} \,
\frac{d\alpha}{dt} = J^{-1} \, \beta_{\alpha} \ .
\label{betaap}
\eeq
with the expansion 
\beq
\beta_{\alpha'} = -2\alpha' \sum_{\ell=1}^\infty b_\ell' (a')^\ell =
-2\alpha' \sum_{\ell=1}^\infty \bar b_\ell' (\alpha')^\ell \ ,
\label{betaprime}
\eeq
where $\bar b'_\ell = b'_\ell/(4\pi)^\ell$. One can then solve for the
$b_\ell'$ in terms of the $b_\ell$ and $k_s$. This yields the known results
that $b_1'=b_1$ and $b_2'=b_2$ \cite{gross75}, and the new results for
$b_\ell'$ at higher loop order $\ell$ that were presented in
\cite{sch}.  Since we will use these higher-loop results for our present work,
we give a relevant list of them in Appendix \ref{bellprime_general}.  It should
be noted that the scheme-independence of $b_2$ assumes that $f(a')$ is
gauge-invariant. This is evident from the fact that in the momentum subtraction
(MOM) scheme, $b_2$ is actually gauge-dependent \cite{mom} and is not equal to
$b_2$ in the $\overline{MS}$ scheme.  We restrict our analysis here to
gauge-invariant scheme transformations and to schemes, such as $\overline{MS}$,
where $b_2$ is gauge-invariant.

The $n$-loop beta function in the transformed scheme, $\beta_{\alpha',n\ell}$,
is given by Eq. (\ref{betaprime}) with the upper limit on the $\ell$ summation
equal to $n$ rather than $\infty$. It is also convenient to define two reduced
beta functions with respective quadratic prefactors extracted, as in our
earlier work, namely
\beq
\beta_{\alpha,n\ell,r} \equiv -\frac{\beta_{\alpha,n\ell,r}}
{2\alpha^2} = \sum_{\ell=1}^n \bar b_\ell \, \alpha^{\ell-1} = 
\frac{1}{4\pi} \sum_{\ell=1}^n b_\ell \, a^{\ell-1} 
\label{beta_nloop_reduced}
\eeq
and similarly 
\beq
\beta_{\alpha',n\ell,r} \equiv -\frac{\beta_{\alpha',n\ell,r}}
{2\alpha'^2} = \sum_{\ell=1}^n \bar b_\ell' \, (\alpha')^{\ell-1} = 
\frac{1}{4\pi} \sum_{\ell=1}^n b_\ell' \, (a')^{\ell-1}  \ . 
\label{betaprime_nloop_reduced}
\eeq

In order to be physically acceptable, a scheme transformation must satisfy
several necessary conditions \cite{sch}.  The first (denoted $C_1$) is that the
scheme transformation must map a real positive $\alpha$ to a real positive
$\alpha'$, since a map taking $\alpha > 0$ to $\alpha'=0$ would be singular,
and a map taking $\alpha > 0$ to a negative or complex $\alpha'$ would
generically violate the unitarity of the theory. Secondly, as condition $C_2$,
the scheme transformation should not map a moderate value of $\alpha$, for
which perturbation theory may be reliable, to a value of $\alpha'$ that is so
large that perturbation theory is unreliable. Thirdly, as condition $C_3$, the
jacobian $J$ should not vanish in the region of $\alpha$ and $\alpha'$ of
interest, or else there would be a pole in Eq. (\ref{betaap}).  The existence
of an IR zero of $\beta$ is a scheme-independent property of an AF theory,
depending (insofar as perturbation theory is reliable) only on the condition
that $b_2 < 0$.  Therefore, as the fourth condition, $C_4$, a scheme
transformation should satisfy the property that $\beta_\alpha$ has an IR zero
if and only if $\beta_{\alpha'}$ has an IR zero. Clearly, these conditions
apply for both a scheme transformation and its inverse.  The conditions can
easily be satisfied by scheme transformations applied in the vicinity of a UV
fixed point at small $\alpha$, but they are not automatically satisfied, and
are a significant restriction, on a scheme transformation applied in the
vicinity of a generic IR fixed point.

% =====================================================================

\section{Scheme Transformations $S_{R,m}$ and $S_{R,\infty}$} 
\label{srm}

In approaching the task of constructing a scheme transformation that maps an
arbitrary initial scheme to the 't Hooft scheme, it is natural to begin by
constructing a family of transformations such that the first removes the
three-loop term in $\beta_{\alpha'}$, i.e., renders $b_3'=0$, the next renders
$b_\ell'=0$ for $\ell=3$ and $\ell=4$, and so forth.  We thus define a scheme
transformation $S_{R,m}$ with $s_{max}=m$ and $m \ge 2$ with the property that
it removes terms in $\beta_{\alpha'}$ from loop order 3 to loop order $m+1$,
inclusive.  That is, starting from an arbitrary initial scheme and applying the
scheme transformation $S_{R,m}$, one has, for the coefficients in the 
transformed scheme, 
\beq
S_{R,m} \ \Longrightarrow \quad  b_\ell'=0 \quad {\rm for} \
\ell=3,...,m+1 \ .
\label{srmbell}
\eeq
Equivalently, $S_{R,m}$ produces the $n$-loop beta function 
$\beta_{\alpha',n\ell}$ in the transformed scheme 
\beqs
\beta_{\alpha',n\ell} & = & -8\pi(a')^2 \bigg [ b_1 + b_2 a' 
+ \sum_{\ell=m+2}^n b'_\ell (a')^{\ell-1} \bigg ] \cr\cr
& = & -2(\alpha')^2 \bigg [ \bar b_1 + \bar b_2 \alpha'
+ \sum_{\ell=m+2}^n \bar b'_\ell (\alpha')^{\ell-1} \bigg ] \ . \cr\cr
& & 
\label{betaprime_nloop_srm}
\eeqs
and the full beta function $\beta_{\alpha'} \equiv \lim_{n \to \infty}
\beta_{\alpha',n\ell}$.  In Eq. (\ref{betaprime_nloop_srm}) and in analogous
equations below, it is understood implicitly that if $n < m+2$, the terms
involving sums over loop order from $\ell=m+2$ to $\ell=n$ are to be replaced
by zero.

There is a unique scheme transformation $S_{R,m}$ that satisfies the properties
that (i) $b'_\ell=0$ for $\ell=3,...,m+1$; (ii) it has unique solutions for all
of the $m$ coefficients $k_s$, $s=1,...,m$, which, in turn, means that these
coefficients are solutions of linear equations. The construction of this scheme
uses the fact that the coefficient $b'_\ell$ for $\ell \ge 3$ contains only a
linear term in $k_{\ell-1}$, so that the equation $b'_\ell=0$ is a linear
equation for $k_{\ell-1}$, which can always be solved uniquely.  To
construct $S_{R,m}$, we take $k_1=0$. Using Eq. (\ref{b3prime}) and solving the
equation $b'_3=0$ for $k_2$, we obtain $k_2 = b_3/b_1$, so
\beq
k_2 = \frac{b_3}{b_1} \quad {\rm for} \ S_{R,m} \ {\rm with} \ m \ge 2 \ . 
\label{k2sol}
\eeq
If we only want to construct $S_{R,2}$, removing the three-loop term in
$\beta_{\alpha'}$, this suffices. If we want to construct $S_{R,m}$ with $m \ge
3$, removing at least the three-loop and four-loop terms in $\beta_{\alpha'}$,
then we need to calculate $k_3$. To do this, we substitute these values of
$k_1$ and $k_2$ into the expression in Eq. (\ref{b4prime}) for $b_4'$ and solve
the equation $b'_4=0$ for $k_3$, obtaining
\beq
k_3 = \frac{b_4}{2b_1} \quad {\rm for} \ S_{R,m} \ {\rm with} \ m \ge 3 \ . 
\label{k3sol}
\eeq
To calculate the coefficient $k_4$ needed
for $S_{R,m}$ with $m \ge 4$, we substitute the above values of $k_s$ 
with $s=1, \ 2, \ 3$ into the expression in Eq. (\ref{b5prime}) for 
$b_5'$ and solve the equation $b_5'=0$ for $k_4$. From this we find that

\beq
k_4 = \frac{b_5}{3b_1} - \frac{b_2b_4}{6b_1^2} 
   + \frac{5b_3^2}{3b_1^2} \quad {\rm for} \ S_{R,m} \ {\rm with} \ m \ge 4 \ .
\label{k4sol}
\eeq

To construct $S_{R,m}$ for higher $m$, we continue iteratively in this manner.
With the set of $k_s$ coefficients calculated up to order $s=m-1$, we calculate
$k_m$ by substituting the solutions for $k_s$, $s=1,...,m-1$, into our
expression for $b_{m+1}'$, then set $b_{m+1}'=0$, and solve for $k_m$.  We list
the resultant $k_s$ for $s=5, \ 6, \ 7$ in Appendix \ref{ks_srm}.  As is clear
from this procedure and from the property that $S_{R,m}$ involves coefficients
$k_s$ with $s=2,...,m$, the explicit construction of the scheme transformation
$S_{R,m}$ in terms of the $b_\ell$ coefficients of the $\beta_\alpha$ function
in an initial scheme requires a knowledge of the $b_\ell$ in this initial
scheme up to the loop order $\ell=m+1$.  Since $s_{max}=m$ for $S_{R,m}$,
\beq
k_s = 0 \quad {\rm for} \ S_{R,m} \ {\rm if} \ s > m \ .
\label{kszero}
\eeq
Using the set of coefficients $k_s$ with $k_1=0$ and $k_s$, $s=2,...,m$ as
calculated in Eqs. (\ref{k2sol}), (\ref{k3sol}), (\ref{k4sol}) and iteratively
for higher $m$, we define the transformation function $f(a')$ for the scheme
transformation $S_{R,m}$:
\beq
f(a')_{S_{R,m}} = 1 + \sum_{s=2}^m k_s (a')^s  \ . 
\label{faprime_srm}
\eeq
Applying this to an initial scheme, we obtain $b_\ell'=0$ for $\ell=3,...,m+1$,
as in (\ref{srmbell})-(\ref{betaprime_nloop_srm}). 

 Some remarks on structural properties of these $k_s$ coefficients are in
order. The coefficient $k_s$ depends on the $b_\ell$ with $\ell=1,...,s+1$ via
the ratios
\beq
\frac{b_\ell}{b_1} \ , \quad {\rm for} \ \ell=2,...,s+1 \ .
\label{belloverb1}
\eeq
It follows that these $k_s$ have the property 
\beq
k_s \ {\rm is \ invariant \ under \ the \ rescaling} \quad b_\ell \to \lambda
b_\ell \ ,
\label{ksinvariance}
\eeq
where $\lambda \in {\mathbb R}$.  A corollary is that
\beq
S_{R,m} \ {\rm is \ invariant \ under \ the \ rescaling} \quad b_\ell \to
\lambda b_\ell \ . 
\label{srninvariance}
\eeq

Since $S_{R,m}$ requires knowledge of the $b_\ell$ up to loop order $\ell=m+1$
and since the $b_\ell$ have been calculated up to $\ell=4$ loops for a general
non-Abelian gauge theory \cite{b3,b4}, it follows that the highest order for
which we can calculate and apply the $S_{R,m}$ scheme transformation is $m=3$.

A scheme transformation that can map an arbitrary initial scheme to a scheme in
which the beta function consists only of the one-loop and two-loop terms
necessarily has $s_{max}=\infty$, since it must remove $m$-loop coefficients up
to arbitrarily high order.  We define $S_{R,\infty} = \lim_{m \to \infty}
S_{R,m} \equiv S_H$. The transformation $S_{R,\infty}$ fulfills the purpose of
mapping an arbitrary initial scheme to a scheme in which $b_\ell'=0$ for all
$\ell \ge 3$, so that the resultant beta function is reduced to just the
(scheme-independent) one-loop and two-loop terms, i.e.,
\beq
S_{R,\infty} \ \Longrightarrow 
\quad \beta_{\alpha'} = -8\pi a^2(b_1 + b_2 a) = 
                  -2\alpha^2(\bar b_1 + \bar b_2 \alpha) \ . 
\label{sh}
\eeq

Since the application of the scheme transformation $S_{R,m}$ to an arbitrary
initial scheme produces a $\beta_{\alpha'}$ function with $b_\ell'=0$ for
$\ell=3,...,m+1$, as expressed in
Eqs. (\ref{srmbell})-(\ref{betaprime_nloop_srm}), it follows that in the new
scheme, the IR zero of the $n$-loop beta function $\beta_{\alpha',m\ell}$ is at
the same value as the (scheme-independent) value $\alpha_{IR,2\ell}$ for
$n$ up to and including $n=m+1$, i.e.,
\beq
S_{R,n} \ \Longrightarrow \ \ 
\alpha_{IR,n\ell}' = \alpha_{IR,2\ell} \quad {\rm for} \ n=3,...,m+1 \ . 
\label{alfir_thesame}
\eeq
%

% =====================================================================

\section{Coefficients $b_\ell'$ Resulting from $S_{R,m}$ Scheme Transformation}
\label{bpsrm}

\subsection{$S_{R,2}$}
\label{bpsr2}

For our applications, it will be useful to exhibit the explicit results for the
coefficients $b'_\ell$ resulting from the applications of the scheme
transformations $S_{R,m}$ with $m=2, \ 3, \ 4$.  In this subsection we show
these for the case $m=2$. Substituting the relevant $k_s$ for the $S_{R,2}$
scheme in the general expressions for the $b'_\ell$, we find
\beq
b'_3=0
\label{bp3_sr2} \ , 
\eeq
\beq
b'_4=b_4 \ , 
\label{bp4_sr2}
\eeq
\beq
b'_5=b_5 + \frac{5b_3^2}{b_1} \ , 
\label{bp5_sr2}
\eeq
\beq
b'_6 = b_6+\frac{2b_3b_4}{b_1}+\frac{3b_2b_3^2}{b_1^2} \ , 
\label{bp6_sr2}
\eeq
\beq
b'_7 = b_7+\frac{3b_3b_5}{b_1}-\frac{9b_3^3}{b_1^2} \ , 
\label{bp7_sr2}
\eeq
\beq
b'_8 = b_8+\frac{4b_3b_6}{b_1}+\frac{4b_3^2b_4}{b_1^2}-\frac{8b_2b_3^3}{b_1^3} 
\ . 
\label{bp8_sr2}
\eeq
In general, after the $S_{R,2}$ scheme transformation is applied, the resultant
$n$-loop beta function, $\beta_{\alpha',n\ell}$, has the form of Eq. 
(\ref{betaprime_nloop_srm}) with $m=2$.

\subsection{$S_{R,3}$}
\label{bpsr3}

From the expressions for $k_s$ in the $S_{R,3}$ scheme transformation, we
calculate the resultant $b'_\ell$ coefficients.  We obtain
\beq
b'_3 = b'_4 = 0 \ , 
\label{bp34_sr3}
\eeq
\beq
b'_5=b_5+\frac{5b_3^2}{b_1}-\frac{b_2b_4}{2b_1} \ , 
\label{bp5_sr3}
\eeq
\beq
b'_6 = b_6+\frac{8b_3b_4}{b_1}+\frac{3b_2b_3^2}{b_1^2} \ , 
\label{bp6_sr3}
\eeq
\beq
b'_7 = b_7+\frac{3b_3b_5}{b_1}+\frac{11b_4^2}{4b_1}
-\frac{9b_3^3}{b_1^2}+\frac{9b_2b_3b_4}{2b_1^2} \ , 
\label{bp7_sr3}
\eeq
\beq
b'_8 = b_8+\frac{4b_3b_6}{b_1}+\frac{b_4b_5}{b_1}-\frac{18b_3^2b_4}{b_1^2}
+\frac{7b_2b_4^2}{4b_1^2}-\frac{8b_2b_3^3}{b_1^3} \ . 
\label{bp8_sr3}
\eeq
After the $S_{R,3}$ scheme transformation is applied, $\beta_{\alpha',n\ell}$
has the form of Eq. (\ref{betaprime_nloop_srm}) with $m=3$.  

We give the corresponding results for the coefficients $b'_\ell$ resulting from
the scheme transformation $S_{R,4}$ in Appendix \ref{bellprime_sr4}.

% ==========================================================================

\section{Application of the $S_{R,2}$ Scheme Transformation}
\label{sr2_application}

As a foundation for our analysis of the scheme transformation $S_{R,3}$, we
recall our results from \cite{sch} concerning $S_{R,2}$ (denoted as $S_2$ in
\cite{sch}).  Let us consider an asymptotically free gauge theory with gauge
group $G$ and $N_f$ massless fermions in a representation $R$ of $G$. Since
\cite{b1,casimir}
\beq
b_1=\frac{1}{3}(11C_A-4T_fN_f) \ , 
\label{b1}
\eeq
the property of asymptotic freedom implies that $N_f < N_{f,b1z}$, where
\cite{nfreal} 
\beq
N_{f,b1z}=\frac{11C_A}{4T_f} \ .
\label{nfb1z}
\eeq
The two-loop coefficient is \cite{b2,jones}
\beq
b_2=\frac{1}{3}\Big [ 34C_A^2-4(5C_A+3C_f)T_f N_f \Big ] \ ,
\label{b2}
\eeq
which decreases monotonically with increasing $N_f$ and
reverses sign as $N_f$ increases through $N_{f,b2z}$, where
\beq
N_{f,b2z} = \frac{34C_A^2}{4(5C_A+3C_f)T_f} \ .
\label{nfb2z}
\eeq
Now for arbitrary $G$ and $R$, 
\beq
N_{f,b2z} < N_{f,b1z} \ ,
\label{nfb2zlessthannfb1z}
\eeq
as is evident from the fact that the difference, 
\beq
N_{f,b1z} - N_{f,b2z} = \frac{3C_A(11C_f+7C_A)}{4T_f(3C_f + 5C_A)} >  0 \ . 
\label{nfdif}
\eeq
Hence, there is always an interval in $N_f$ such that $b_1 > 0$ while $b_2 <
0$, so that the two-loop ($2\ell$) beta function has an IR zero.  We denote
this interval as $I$:
\beq
I: \quad N_{f,b1z} < N_f < N_{f,b2z} \ .
\label{nfinterval}
\eeq
The zero of the two-loop beta function (which is scheme-independent) occurs at
$\alpha=\alpha_{IR,2\ell}$, where 
\beq
\alpha_{IR,2\ell} = -\frac{4\pi b_1}{b_2} 
\label{alfir_2loop}
\eeq
(i.e., $a_{IR,2\ell}=-b_1/b_2$), which is physical for $N_f \in I$. 
From the $m=2$ special case of Eq. (\ref{alfir_thesame}), it follows that after
the application of the $S_{R,2}$ scheme transformation, in terms of the new
variable $\alpha'$, 
\beq
\alpha'_{IR,3\ell}=\alpha'_{IR,2\ell}=\alpha_{IR,2\ell} \ .
\label{alfthesame_sr2}
\eeq

For the $S_{R,2}$ scheme transformation, the function $f(a')$ has the
form 
\beq 
S_{R,2}: \quad f(a') = 1 + \frac{b_3}{b_1}(a')^2 = 1 +
\frac{\bar b_3}{\bar b_1} (\alpha')^2 \ .
\label{fap_sr2}
\eeq

Now we assume that $N_f \in I$, so that there is an IR zero in $\beta_{2\ell}$,
as given in Eq. (\ref{alfir_2loop}).  We start in the $\overline{\rm SM}$
scheme and operate with the $S_{R,2}$ scheme transformation. We evaluate
$f(a')$ at this IR zero, $a'_{IR,2\ell}=a_{IR,2\ell}=-b_1/b_2$ and
obtain the following result: 
\beq
S_{R,2}: \quad f(a'_{IR,2\ell}) = 1 + \frac{b_1b_3}{b_2^2} 
                                = 1 + \frac{\bar b_1 \bar b_3}{\bar b_2^2} \ . 
\label{fap_sr2_air2loop}
\eeq
In order that this transformation obey condition $C_1$, namely that it maps $a'
> 0$ to $a > 0$, we require that $f(a') > 0$. This inequality must be
satisfied, in particular, at $a'_{IR,2\ell}=a_{IR,2\ell}$, so we obtain the
inequality
\beq
1 + \frac{\bar b_1 \bar b_3}{\bar b_2^2} >  0 \ .
\label{fap_sr2_inequality}
\eeq
Since $\bar b_3 < 0$ for $N_f \in I$ in the $\overline{\rm MS}$ scheme, and,
more generally, in schemes that maintain at the three-loop level the IR zero in
the two-loop beta function \cite{bc}, we can also write this in terms of
positive quantities as the condition that 
\beq
1 - \frac{\bar b_1 |\bar b_3|}{\bar b_2^2} >  0 \ .
\label{fap_sr2_inequality2}
\eeq
This analysis holds for an arbitrary gauge group $G$ and fermion content such
that the two-loop $\beta$ function has an IR zero.  

As was shown in \cite{sch}, the inequality (\ref{fap_sr2_inequality}) is not,
in general, satisfied, so this $S_{R,2}$ scheme transformation violates
condition $C_1$ in the vicinity of the IR fixed point for a certain range of
smaller values of $N_f \in I$. To show the violation of the inequality
(\ref{fap_sr2_inequality}), it suffices to consider the class of theories with
$G={\rm SU}(N_c)$ and $N_f$ fermions in the fundamental representation.  The
interval $I$ where the two-loop $\beta$ function has an IR zero is then
\beq
I: \quad \frac{34N_c^3}{13N_c^2-3} < N_f < \frac{11N_c}{2} \ . 
\label{nfrange}
\eeq
For $N_c=2$, the interval $I$ is $5.55 < N_f < 11$, while for For $N_c=3$, $I$
is $8.05 < N_f < 16.5$.  For $N_c=2$, the inequality (\ref{fap_sr2_inequality})
is violated for $5.55 < N_f < 8.44$ and is satisfied for $8.44 < N_f < 11$,
while for $N_c=3$, inequality is violated for $8.05 < N_f < 12.41$ and is
satisfied for $12.41 < N_f < 16.5$.  Note that the same is true if $f(a')$ is
evaluated for $a'=a'_{IR,3\ell}$, since by Eq. (\ref{alfir_thesame}),
$a'_{IR,3\ell}=a'_{IR,2\ell}=a_{IR,2\ell}$.  In Table I of \cite{sch} we gave
the values of $\alpha'_{IR,4\ell}$ resulting from the application of the
$S_{R,2}$ scheme transformation. In Table \ref{fapvalues} of the present paper
we list the values of $f(a'_{IR,2\ell})$ for this $S_{R,2}$ scheme
transformation, for the illustrative values $2 \le N_c \le 4$ and $N_f$ in the
respective intervals $I$ for each $N_c$. Given a value of $N_c$, for values of
$N_f$ near the lower end of the respective interval $I$, $|f(a')|$ gets large
compared to unity.  This is a consequence of the fact that $b_2 \to 0$ at this
lower boundary of the interval $I$ and hence formally, $\alpha_{IR,2\ell}$
diverges.  Thus, for these values of $N_f$, in addition to the fact that this
$S_{R,2}$ scheme transformation violates condition $C_1$ because $f(a')$ is
negative, it also violates condition $C_4$, because it maps moderate values of
the gauge coupling to values that are too large for perturbation theory to be
reliable.

% =========================================================================

\section{Application of the $S_{R,3}$ Scheme Transformation}
\label{sr3_application}

We next address and answer the question of whether one can alleviate or
circumvent the pathology encountered with $S_{R,2}$ at an IR fixed point
(negative $f(a')$ for various $N_f \in I$) by instead using $S_{R,3}$.  The
transformation function $f(a')$ for $S_{R,3}$ is
\beqs
S_{R,3}: \ f(a') & = & 1 + k_2 (a')^2 + k_3 (a')^3 \cr\cr
            & = & 1 + \frac{b_3}{b_1} (a')^2 + \frac{b_4}{2b_1} (a')^3 \cr\cr
            & = & 1 + \frac{\bar b_3}{\bar b_1} (\alpha')^2 + 
                 \frac{\bar b_4}{2\bar b_1} (\alpha')^3 \ . 
\label{fap_sr3}
\eeqs
From the $m=3$ special case of Eq. (\ref{alfir_thesame}), it follows that after
the application of the $S_{R,3}$ scheme transformation, in terms of the new
variable $\alpha'$, 
\beq
\alpha'_{IR,4\ell}=\alpha'_{IR,3\ell}=\alpha'_{IR,2\ell}=\alpha_{IR,2\ell} \ .
\label{alfthesame_sr3}
\eeq

We use the same technique as for the analysis of $S_{R,2}$, namely we consider
$N_f \in I$, so that $\beta_{2\ell}$ has an IR zero. Evaluating $f(a')$ at this
(scheme-independent) two-loop zero,
$a'_{IR,2\ell}=a_{IR,2\ell}=-b_1/b_2$, we have
\beqs
f(a'_{IR,2\ell}) & = & 1 + \frac{b_1 b_3}{b_2^2} - \frac{b_1^2 \, b_4}{2b_2^3} 
\cr\cr
& = & 1 + \frac{\bar b_1 \bar b_3}{\bar b_2^2} 
- \frac{\bar b_1^2 \, \bar b_4}{2\bar b_2^3} \ . 
\label{fap_sr3_air2loop}
\eeqs
In order for the $S_{R,3}$ scheme transformation to be acceptable, a necessary
condition is $C_1$, that $f(a') > 0$, in particular, at 
$a'=a_{IR,2\ell}=a_{IR,2\ell}$, i.e., that 
\beq
1 + \frac{\bar b_1 \bar b_3}{\bar b_2^2} 
- \frac{\bar b_1^2 \, \bar b_4}{2 \bar b_2^3} > 0 \ . 
\label{fap_sr3_inequality}
\eeq
Now $b_2 < 0$ for $N_f \in I$ and, as shown in \cite{bc} $b_3 < 0$ for $N_f \in
I$ not only in the $\overline{\rm MS}$ scheme, but more generally in any scheme
that has the necessary property of maintining the scheme-independent property
that the two-loop $\beta$ function has an IR zero.  Given these properties, we
can reexpress (\ref{fap_sr3_inequality}) in terms of positive quantities as
\beq
1 - \frac{\bar b_1 |\bar b_3|}{\bar b_2^2} 
+ \frac{\bar b_1^2 \, \bar b_4}{2|\bar b_2|^3} > 0 \ . 
\label{fap_sr3_inequality2}
\eeq
As is evident in Table I of \cite{bvh}, for $N_c=2$ and $N_c=3$, $b_4$ is
positive for all $N_f$ in the respective intervals $I$, but for $N_c \ge 4$,
$b_4$ can be negative for some value(s) of $N_f \in I$. 

This analysis for $S_{R,3}$ holds for an arbitrary gauge group $G$ and fermion
representation such that $N_f \in I$.  For our present purposes, it will
suffice to consider the case $G={\rm SU}(N_c)$ and fermions in the fundamental
representation.  As before, we start in the $\overline{\rm MS}$ scheme.  In
Table \ref{fapvalues} we list values of $f(a'_{IR,2\ell})$ for the $S_{R,3}$
scheme transformation, for $N_c=2, \ 3, \ 4$ and $N_f$ in the respective
intervals $I$ for each $N_c$ where the two-loop beta function has an IR zero.
As we noted above in the case of $S_{R,2}$, for smaller values of $N_f$ 
in the respective interval $I$ for each $N_c$, $|f(a')|$ is substantially
larger than unity, so that, in addition to the violation of condition $C_1$, 
$f(a')$ also violates condition $C_4$.  We omit entries for the lowest values 
of $N_f$ in the respective intervals $I$, for which $|f(a')| >> 1$, where this
violation is most extreme; for example, for SU(3) with $N_f=9$, 
$f(a')_{IR,S_{R,2}}=-19.851$ and $f(a')_{IR,S_{R,3}}=-15.282$. 

From our Table \ref{fapvalues}, one sees that for $N_c=2, \ 3, \ 4$, the values
of $N_f$ that yield an unphysical negative $f(a')$ for the $S_{R,2}$ scheme
transformation also yield an unphysical negative $f(a')$ for the $S_{R,3}$
scheme transformation.  This is also true for almost all of the values of $N_f$
in the case $N_c=5$, with one exception; for $N_f=20$, $f(a_{IR,2\ell}) < 0$
for $S_{R,2}$, while $f(a_{IR,2\ell}) > 0$ for $S_{R,3}$.  We have also
investigated this for higher $N_c$, with similar findings. We note that the
same results are obtained by substituting the three-loop IR zero, since by
Eq. (\ref{alfir_thesame}) this is equal to the scheme-independent two-loop IR
zero.  Therefore, using the $S_{R,3}$ scheme transformation does not alleviate
the problem encountered with the $S_{R,2}$ transformation and does not
significantly increase the range of $N_f$ where $f(a')$ satisfies the necessary
condition of being positive when evaluated at the scheme-independent value
$a_{IR,2\ell}$. These $S_{R,m}$ scheme transformations can be used for larger
values of $N_f$ toward the upper end of the interval $I$, where
$\alpha_{IR,2\ell}$ is correspondingly smaller, approaching zero as $N_f
\nearrow N_{f,b1z}$.  However, to show this for $S_{R,\infty}$ is delicate,
since it requires that one analyze the convergence of an infinite series
\cite{sch}.

In passing, we remark on a related topic.  For this purpose, let us consider a
general (vectorial) gauge theory with an arbitrary non-Abelian gauge group $G$
and $N_f$ fermions in an arbitrary representation.  For a given $G$ and $R$,
let us consider increasing $N_f$ past the value where $b_1$ reverses sign, so
that the theory becomes non-asymptotically free.  One may ask whether this
theory has an ultraviolet fixed point and if so, what is the range of
applicability of the $S_{R,m}$ scheme transformation for various $m$. Because
of the inequality (\ref{nfb2zlessthannfb1z}), it follows that if $N_f >
N_{f,b1z}$ (so $b_1 < 0$), then also $N_f > N_{f,b2z}$, so that $b_2 <
0$. Hence, this theory has no two-loop zero in its $\beta$ function.  Since
this is the maximal scheme-independent information that one has, even if one
were to obtain a zero of the $\beta$ function at higher loops (which would now
be a UV fixed point), one could not convincingly argue that this is physical.
Below we shall discuss this sort of question further for a UV zero in the
$\beta$ function of a U(1) gauge theory.

% =========================================================================

\section{$S_{R,m}$ Scheme Transformation in the Limit $N_c \to \infty$, 
$N_f \to \infty$ with $N_f/N_c$ fixed}
\label{lnn_section}

For the case of $G={\rm SU}(N_c)$ and $N_f$ fermions in the fundamental
representation, a limit of particular interest is the 't Hooft-Veneziano limit
\cite{thooftveneziano},
\beq
N_c \to \infty, \quad N_f \to \infty, \quad {\rm with} \ \ 
r \equiv \frac{N_f}{N_c} \ \ {\rm fixed} \ .
\label{lnn}
\eeq
In this limit, one also requires that the product
\beq
\xi(\mu) \equiv \alpha(\mu)N_c 
\label{xi}
\eeq
be a fixed, finite function of $\mu$. We denote this as the LNN (large $N_c$
and $N_f$) limit.

Here we investigate the applicability of the $S_{R,2}$ and $S_{R,3}$ scheme
transformations for $N_f \in I$ in the LNN limit. One of the reasons for the
interest in the LNN limit is that properties of the $\beta$ function exhibit an
approximate universality, in the sense that they are similar for different
values of $N_c$ and $N_f$ if the ratio $r=N_f/N_c$ is similar or the same
\cite{bvh,bc}.  The study in \cite{lnn} gave some insight into the origin of
this universality.

To construct an appropriate beta function that has a finite, nontrivial LNN
limit, one multiplies both sides of Eq. (\ref{beta}) by $N_c$ and then takes
this limit, obtaining a result that is a function of $\xi$,
\beq
\beta_{\xi} \equiv \frac{d\xi}{dt} = \lim_{LNN} \beta_\alpha N_c \ .
\label{betaxi}
\eeq
This beta function has the expansion
\beq
\beta_\xi \equiv \frac{d\xi}{dt}
= -8\pi x \sum_{\ell=1}^\infty \hat b_\ell x^\ell
= -2 \xi \sum_{\ell=1}^\infty \tilde b_\ell \xi^\ell \ ,
\label{betaxiseries}
\eeq
where $x = \xi/(4\pi)$ and
\beq
  \hat b_\ell = \lim_{LNN} \frac{b_\ell}{N_c^\ell} \ , \quad
\tilde b_\ell = \lim_{LNN} \frac{\bar b_\ell}{N_c^\ell} \ .
\label{bellrel}
\eeq
Thus, $\tilde b_\ell = \hat b_\ell/(4\pi)^\ell$. One defines the $n$-loop
$\beta_\xi$ function by Eq. (\ref{betaxiseries}) with the upper limit on the 
summation over loop order $\ell=\infty$ replace by $\ell=n$.

The (scheme-independent) one-loop and two-loop coefficients in $\beta_\xi$ are
\beq
\hat b_1 = \frac{1}{3}(11-2r)
\label{b1hat}
\eeq
and
\beq
\hat b_2 = \frac{1}{3}(34-13r)  \ .
\label{b2hat}
\eeq
Asymptotic freedom requires that $b_1 > 0$ and hence that $r < 11/2$. The
coefficient $\hat b_2$ reverses sign to negative values as $r$ increases
through the value $r=34/13$.  Consequently, for $r$ in the real interval 
\beq
I_r \ \frac{34}{13} < r < \frac{11}{2} \ , 
\label{rinterval}
\eeq
i.e., $2.6154 < r < 5.5$, $\beta_{\xi,2\ell}$ has an IR zero.  This zero occurs
at
\beq
\xi_{IR,2\ell} = 4\pi x_{IR,2\ell} = \frac{4\pi(11-2r)}{13r-34} \ . 
\label{xiir_2loop}
\eeq
The three-loop and four-loop coefficients $\hat b_3$ and $\hat b_4$ were given
in \cite{lnn}.

A scheme transformation applicable to the theory in the LNN limit is thus 
\beq
x = x'f(x') \ . 
\label{xxp}
\eeq
One requires that $f(0)=1$ to keep the UV properties the same.
Considering $f(x')$ that are analytic at $x'=x=0$, one has the expansion
\beq
f(x') = 1 + \sum_{s=1}^{s_{max}} k_s (x')^s =
        1 + \sum_{s=1}^{s_{max}} \bar k_s (\xi')^s \ . 
\label{fxprime}
\eeq

Evaluating the $S_{R,2}$ expression for 
$f(x')$ in the LNN limit at $x=x_{IR,2\ell}$, we calculate 
\beqs
& & S_{R,2; LNN} \ \Longrightarrow \cr\cr
& &  f(x')_{IR,2\ell} = 
\frac{52235 - 40425r + 7692r^2 - 224r^3}{18(13r-34)^2} \ . \cr\cr
& & 
\label{fap_sr2_lnn}
\eeqs
For $r \in I_r$, this $f(x')$ is a monotonically increasing function of $r$,
which passes through zero from negative to positive values as $r$ increases
through the value $r = 4.06814$ (quoted to the indicated accuracy). Thus, 
\beqs
S_{R,2; LNN} \ \Longrightarrow & & 
    f(x')_{IR,2\ell} < 0 \quad {\rm for} \ 2.6154 < r < 4.0681 \cr\cr
& & f(x')_{IR,2\ell} > 0 \quad {\rm for} \ 4.0681 < r < 5.5000 \ . \cr\cr
& & 
\label{fap_sr2_lnn_sign}
\eeqs

Evaluating the $S_{R,3}$ expression for $f(x')$ in the LNN limit at
$x=x_{IR,2\ell}$, we obtain
\begin{widetext}
\beqs
S_{R,3; LNN} \Longrightarrow f(x')_{IR,2\ell} & = & 
\frac{1}{6^4(13r-34)^3} \, 
\bigg [ -55042348 + 62622039r - 24520604r^2 + 2885644r^3 + 21504r^4 
+ 4160r^5 \cr\cr
& + & \zeta(3)\Big ( 1149984 - 940896r + 2423520r^2 - 815616r^3 + 72576r^4 
 \Big ) \bigg ] \ . 
\label{fap_sr3_lnn}
\eeqs
\end{widetext}
Here, $\zeta(s) \equiv \sum_{n=1}^\infty n^{-s}$ is the Riemann zeta function,
with $\zeta(3) = 1.202057$, etc.  For $r \in I_r$, this $f(x')$ is again a
monotonically increasing function of $r$, which passes through zero from
negative to positive values as $r$ increases through the value $r = 3.95069$
(to the indicated accuracy). Thus,
\beqs
S_{R,3; LNN} \ \Longrightarrow & & 
    f(x')_{IR,2\ell} < 0 \quad {\rm for} \ 2.6154 < r < 3.9507 \cr\cr
& & f(x')_{IR,2\ell} > 0 \quad {\rm for} \ 3.9507 < r < 5.5000 \ . \cr\cr
& & 
\label{fap_sr3_lnn_sign}
\eeqs
Evidently, the positivity properties of $f(x')$ for the $S_{R,2}$ and
$S_{R,3}$ scheme transformations are quite similar in this LNN limit.  This is
in agreement with our calculations in Table \ref{fapvalues} for specific values
of $N_c$ and $N_f$.  Clearly, in the respective intervals of $r$ where $f(x')$
is negative, the $S_{R,2}$ and $S_{R,3}$ scheme transformations are
unacceptable, since they fail to satisfy the condition $C_1$. 

% ========================================================================

\section{ U(1) Gauge Theory}
\label{u1} 

\subsection{General} 
\label{u1general}

It is also of interest to explore the effects of higher-order terms and the
associated scheme dependence in the $\beta$ function for an Abelian gauge
theory.  We consider the simplest example of such a theory, namely a vectorial
theory with a U(1) gauge group and $N_f$ fermions of charge $q$. We use the
same notation for the gauge coupling $g(\mu)$ and for $\alpha(\mu)$ and
$a(\mu)=\alpha(\mu)/(4\pi)$ as before.  With no loss of generality, we absorb
$q$ into the definition of $g$.  As is well known, this theory is not
asymptotically free and must be regarded as a low-energy effective field
theory.  One may investigate whether the two-loop $\beta$ function for this
theory has a zero, which would thus be an exact or approximate ultraviolet
fixed point (UVFP).  If, indeed, such a UV zero were present in the $\beta$
function, one could also study the effect of the $S_{R,m}$ scheme
transformation on its value.  In contrast to the case of an IRFP in an
asymptotically free theory, here the UV to IR evolution would be envisioned as
starting from the UVFP and flowing to weaker coupling.

For convenience, we define $\beta_\alpha$ for this theory without the minus
sign prefactor in Eq. (\ref{beta}). The coefficients that have been calculated
can be obtained from those for the non-Abelian theory by the formal
replacements $C_A=0$, $C_f=1$, and $T_f=1$, together with replacements of other
group invariants that enter at the four-loop level \cite{b4}.  If one fixes
$\alpha(\mu)$ at a some high scale $\mu=\Lambda$ in the ultraviolet, then for a
U(1) gauge theory with fermions of negligibly small mass, $\alpha(\mu) \to 0$
as $\mu \to 0$, so the theory becomes free in the infrared (often called the
triviality property).  Actually, because there is no confinement, an U(1)
theory with exactly zero-mass charged fermions has problems with infrared
divergences, so a more precise statement of this property is that for the U(1)
theory with fermions of mass $m_0 << \Lambda$, the running coupling
$\alpha(m_0)$ becomes arbitrarily small as $m_0/\Lambda \to 0$. If one were to
take $\mu < m_0$, then in the construction of the low-energy effective field
theory applicable in this interval, one would integrate out the fermions, and
thereby obtain a free theory.  Viewed the other way, from the IR to the UV, if
one fixes $\alpha(\mu)$ at some scale in the infrared such as $\mu = m_0$ and
then increases $\mu$, a solution of the one-loop $\beta$ function equation
yields a Landau pole. Of course, the perturbative calculation that produces
this result is not reliable when $\alpha(\mu)$ becomes so large as to approach
this pole.  Moreover, this would not be relevant to the actual physics if the
ultraviolet completion of the U(1) gauge theory involves embedding of the U(1)
factor group in an asymptotically free simple non-Abelian gauge group, as is
the case with the embedding of the weak hypercharge U(1)$_Y$ factor group in a
grand unified theory.  It may be recalled that among the motivations for 
grand unification, one is that this embedding provides an elegant explanation
of the quantization of weak hypercharge and hence electric charge in the
Standard Model.  

A number of studies have been performed to investigate the properties of U(1)
gauge theory with fermions using methods going beyond perturbation theory, such
as approximate solutions of Schwinger-Dyson equations \cite{jbw,scgt90} and
simulations of the theory on a lattice \cite{scgt90,schierholz}.  In
particular, fully nonperturbative lattice studies were carried out with
dynamical staggered fermions (effectively corresponding to $N_f=4$ continuum
fermion species) and led to the conclusion that this theory does not have a
(nontrivial) UV fixed point \cite{scgt90,schierholz}.  This question has also
been examined using analytic results for the large-$N_f$ limit of the theory
\cite{largeNf,ps}.  For our present purposes, we focus on the specific question
of the scheme-dependence of a possible UV zero in $\beta$, as is manifested in
the effects of higher-loop terms.  This is timely in part because the five-loop
term in $\beta$ has recently been calculated, as discussed below.

% ======================================================================

\subsection{$\beta_{\alpha,2\ell}$} 
\label{u1alfuv_2loop_section}

Given that for this U(1) gauge theory we define $\beta_\alpha$ as in
Eq. (\ref{beta}) but without the minus sign prefactor, the one-loop and
two-loop coefficients are \cite{rg,b2u1}
\beq
b_1 = \frac{4N_f}{3}
\label{b1u1}
\eeq
and
\beq
b_2 = 4N_f \ . 
\label{b2u1}
\eeq
Because $b_1$ and $b_2$ have the same sign, the two-loop $\beta$ function,
$\beta_{\alpha,2\ell}$, for this U(1) theory does not have a UV zero.  As noted
above, the two-loop beta function embodies the maximal scheme-independent
information on the coupling constant evolution of the theory.  Of course, this
analysis is within the context of the perturbatively calculated $\beta$
function and does not address the possibility of a nonperturbative UV zero.
Owing to the absence of a UV zero in the two-loop $\beta$ function, we cannot
use the same method that we employed above to test the applicability of a
scheme transformation, namely to evaluate $f(a')$ at the two-loop zero and
check to see where it is positive and of moderate size.  Consequently, in order
to study scheme-dependent effects in the context of a possible UV zero in the
$\beta$ function, we will simply investigate whether, for a given $N_f$, the
higher-loop terms in $\beta$ lead to a UV zero, and, if so, how the
location of this zero changes as a function of loop order.  Because of the 
absence of a UV zero in $\beta$ at the two-loop level, even where it is present
at the scheme-dependent higher-loop order, this perturbative analysis does not
yield convincing evidence that it is physical.

% ========================================================================

\subsection{ $\beta_{\alpha,3\ell}$} 
\label{u1alfuv_3loop_section}

In the $\overline{\rm MS}$ scheme, the three-loop coefficient in the $\beta$
function of the U(1) gauge theory has the negative-definite value
\cite{b3u1Nf1,b3u1}
\beq
b_3 = -2N_f\Big ( 1 + \frac{22N_f}{9} \Big ) \ , 
\label{b3u1}
\eeq
so that the three-loop beta function is 
\beq
\beta_{\alpha,3\ell} = 8\pi N_f a^2 \bigg [ \frac{4}{3} +
4 a - 2\Big ( 1 + \frac{22N_f}{9} \Big )a^2 \bigg ] \ . 
\label{beta_alf_u1_3loop}
\eeq
Thus, in addition to the IR zero at $\alpha=0$, in the $\overline{\rm MS}$
scheme, $\beta_{\alpha,3\ell}$ vanishes at the UV zero 

\beq
\alpha_{UV,3\ell}= 4 \pi a_{UV,3\ell} = 
 \frac{4\pi[9 + \sqrt{3(45+44N_f)} \ ]}{9+22N_f}
\label{alfuv_3loop}
\eeq
(and, formally, at an unphysical negative value of $a$ given by the above
expression with a minus sign in front of the square root).  We list values of
$\alpha_{UV,3\ell}$ in Table \ref{alfuvu1} as a function of $N_f$ for $N_f=1$
to $N_f=10$.

From Eq. (\ref{alfuv_3loop}), it follows that, in the ${\overline {\rm MS}}$
scheme, this $\alpha_{UV,3\ell}$ is a monotonically decreasing function of
$N_f$. As $N_f \to \infty$, $\alpha_{UV,3\ell}$ approaches zero like
\beq
\alpha_{UV,3\ell} = 4\pi\sqrt{ \frac{3}{11N_f}} \bigg [ 1 +
\frac{3}{22}\sqrt{\frac{33}{N_f}} + \frac{9}{88N_f} + 
O \Big ( \frac{1}{(N_f)^{3/2}} \Big ) \bigg ] 
\label{alf_uv_3loop_largeNf}
\eeq
Note that, even apart from the scheme-dependence, for moderate $N_f$, the value
of $\alpha_{UV,3\ell}$ in Eq. (\ref{alfuv_3loop}) is too large for
the perturbative three-loop calculation to be very accurate.  The fact that 
$\alpha_{UV,3\ell} \sim O(1)$ means that higher-loop corrections are
generically important.  We turn next to these. 

% ========================================================================

\subsection{ $\beta_{\alpha,4\ell}$} 
\label{u1alfuv_4loop_section}

In the $\overline{\rm MS}$ scheme the four-loop coefficient in the $\beta$
function of the U(1) gauge theory is \cite{b4u1,math} 
\beq
b_4 = N_f\Big [ -46 + \Big ( \frac{760}{27} - \frac{832}{9}\zeta(3) 
\Big ) N_f - \frac{1232}{243}N_f^2 \Big ] \ . 
\label{b4u1}
\eeq
Numerically, 
\beq
b_4 = -N_f \, (46+82.97533N_f+5.06996N_f^2) . 
\label{b4num}
\eeq
Evidently, $b_4 < 0$ for all $N_f > 0$. The condition that
$\beta_{\alpha,4\ell}=0$ for $\alpha \ne 0$, is the cubic equation in $\alpha$,
or equivalently, $a$, \ $b_1+b_2a+b_3a^2+b_4a^3=0$.  This equation has a
physical root, $a_{UV,4\ell}=\alpha_{UV,4\ell}/(4\pi)$, as well as an
unphysical pair of complex-conjugate values of $a$.  We list values of
$\alpha_{UV,4\ell}$ in Table \ref{alfuvu1} as a function of $N_f$.  As was the
case with $\alpha_{UV,3\ell}$, in this $\overline{\rm MS}$ scheme,
$\alpha_{UV,4\ell}$ is a monotonically decreasing function of $N_f$.  We find
that when one goes from three loops to four loops, the UV zero decreases, i.e.,
\beq
\alpha_{UV,4\ell} < \alpha_{UV,3\ell} \quad {\rm for \ fixed} \ N_f \ .
\label{alfuv_34loop_inequality}
\eeq
This decrease is substantial, roughly by a factor of 2. 

% =====================================================================

\subsection{ $\beta_{\alpha,5\ell}$} 
\label{u1alfuv_5loop_section}

Recently, the five-loop coefficient has been calculated to be \cite{b5u1}  
\begin{widetext} 
\beqs
b_5 & = & N_f \bigg [ \frac{4157}{6}+128\zeta(3) 
+ \Big (-\frac{7462}{9}-992\zeta(3)+2720\zeta(5) \Big ) N_f \cr\cr
& + & \Big (-\frac{21758}{81}+\frac{16000}{27}\zeta(3)-\frac{416}{3}\zeta(4)
-\frac{1280}{3}\zeta(5) \Big )N_f^2 + 
\Big (\frac{856}{243}+\frac{128}{27}\zeta(3) \Big )N_f^3 \bigg ] \ . 
\label{b5u1}
\eeqs
\end{widetext}
Numerically, 
\beqs
b_5 & = & N_f( 846.6966 + 798.8919N_f - 148.7919N_f^2 \cr\cr
    & + & 9.22127N_f^3). 
\label{b5u1num}
\eeqs
This is positive for all non-negative $N_f$, both integral and real.  The
condition that $\beta_{\alpha,5\ell}$ vanishes away from the origin is the
quartic equation $\sum_{\ell=1}^5 b_\ell a^{\ell-1}=0$.  We find that for $N_f
= 1, \ 2, \ 3, \ 4$, this equation has no physical solutions. (It has two pairs
of complex-conjugate solutions.) For $N_f \ge 5$, we find that there are two
positive real roots to this equation; the smaller of these is
$a_{UV,5\ell}$. We list the corresponding values of $\alpha_{UV,5\ell}$ in
Table \ref{alfuvu1} as a function of $N_f$. As is evident from this table, for
values of $N_f$ where the theory exhibits a physical value of
$\alpha_{UV,5\ell}$ in this $\overline{\rm MS}$ scheme, it is a monotonically
decreasing function of $N_f$. We find that 
\beq
\alpha_{UV,5\ell} > \alpha_{UV,4\ell} \quad {\rm for \ fixed} \ N_f \ .
\label{alfuv_45loop_inequality}
\eeq
However, the slight increase in the value of the UV zero of $\beta$ going from
four-loop to five-loop order is smaller than the magnitude of the decrease
going from three-loop to four-loop order, so that, for $N_f$ values where
$\beta_{5\ell}$ has a UV zero, 
\beq
\alpha_{UV,5\ell} < \alpha_{UV,3\ell} \quad {\rm for \ fixed} \ N_f \ .
\label{alfuv_35loop_inequality}
\eeq
As is evident from Table \ref{alfuvu1}, $\alpha_{UV,5\ell}$ is approximately
half of the value of $\alpha_{UV,3\ell}$.  These higher-loop results provide a
quantitative measure of the effect of scheme dependence in the $\beta$
function of the U(1) gauge theory. 

% =========================================================================

\section{Conclusions}
\label{conclusions} 

Because terms at loop order $\ell \ge 3$ in the $\beta$ function of a gauge
theory are scheme-dependent, it follows that one can carry out a scheme
transformation to remove these terms at sufficiently small coupling.  A basic
question concerns the range of applicability of such a scheme transformation.
It is particularly important to address this question when studying the IR zero
that is present in the $\beta$ function of an asymptotically free gauge theory
for certain types of fermion content. In this paper, extending the study in
\cite{sch}, we have studied the properties of the scheme transformation
$S_{R,m}$ with $m \ge 2$, which renders the beta function coefficients
$b'_\ell=0$ for $3 \le \ell \le m+1$, at least for sufficiently small
$\alpha$. We have calculated and presented expressions for the nonzero
coefficients $b'_\ell$ with $\ell \ge m+2$ resulting from the application of
the $S_{R,m}$ scheme transformation, up to the loop order $\ell=8$. Since
calculations with the scheme transformation $S_{R,m}$ require a knowledge of
the terms in the $\beta$ function up to loop order $\ell+1$, $S_{R,3}$ is the
highest-order scheme transformation of this type that can be analyzed
explicitly for a general non-Abelian gauge theory, using the beta function
coefficients calculated up to four-loop order.  We have carried out this
analysis and have shown that the range of $N_f$ values where the $S_{R,3}$
scheme transformation is applicable is limited to $N_f$ values in the upper
part of the interval $I$ where the two-loop $\beta$ function has an IR zero at
a correspondingly small value, $\alpha_{IR,2\ell}$.  We have shown that this
range of applicability is similar to that found for the $S_{R,2}$ scheme
transformation. For example, for an SU(3) gauge theory with $N_f=12$ fermions,
neither $S_{R,2}$ nor $S_{R,3}$ can be used to study the IR fixed point because
they produce unphysical effects. Our results elucidate the limitations on the
use of scheme transformations to remove terms at loop order $\ell \ge 3$ in the
beta function of a gauge theory, a subject that does not seem to have received
much attention in the literature.  These results add to one's knowledge of the
UV to IR evolution of an asymptotically free gauge theory, a fundamental topic
in quantum field theory. We have also investigated scheme-dependent effects of
higher-loop terms in the $\beta$ function of a U(1) gauge theory.

% =======================================================================

\begin{acknowledgments}
This research was partially supported by the NSF grant NSF-PHY-09-69739. 
\end{acknowledgments}

% ========================================================================

\begin{appendix}

\section{Equations for the $b_\ell'$ Resulting from a General Scheme
Transformation}
\label{bellprime_general}

The expressions for the $b_\ell'$ in Eq. (\ref{betaprime}) 
for $3 \le \ell \le 6$ are \cite{sch} 
\beq
b_3' = b_3 + k_1b_2+(k_1^2-k_2)b_1 \ , 
\label{b3prime}
\eeq
\beq
b_4' = b_4 + 2k_1b_3+k_1^2b_2+(-2k_1^3+4k_1k_2-2k_3)b_1 \ , 
\label{b4prime}
\eeq
\begin{widetext}
\beq
b_5' = b_5+3k_1b_4+(2k_1^2+k_2)b_3+(-k_1^3+3k_1k_2-k_3)b_2
     + (4k_1^4-11k_1^2k_2+6k_1k_3+4k_2^2-3k_4)b_1 
\label{b5prime}
\eeq
and
\beqs
b_6' & = & b_6 +4k_1b_5+(4k_1^2+2k_2)b_4+4k_1k_2b_3
 + (2k_1^4-6k_1^2k_2+4k_1k_3+3k_2^2-2k_4)b_2 \cr\cr
     & + & (-8k_1^5+28k_1^3k_2-16k_1^2k_3-20k_1k_2^2
      + 8k_1k_4+12k_2k_3 -4k_5)b_1 \ . 
\label{b6prime}
\eeqs
\end{widetext}
%

% =========================================================================

\section{Higher-Order Coefficients for $S_{R,m}$} 
\label{ks_srm}

In this appendix we list expressions for some higher-order coefficients 
$k_s$ in the $S_{R,m}$ scheme transformation.  We calculate that 
\begin{widetext}
\beq
k_5 = \frac{b_6}{4b_1} - \frac{b_2b_5}{6b_1^2} + \frac{2b_3b_4}{b_1^2}
+ \frac{b_2^2b_4}{12b_1^3} - \frac{b_2b_3^2}{12b_1^3} 
\quad {\rm for} \ S_{R,m} \ {\rm with} \ m \ge 5 \ , 
\label{k5thooftsol}
\eeq
\beq
k_6 = \frac{b_7}{5b_1} - \frac{3b_2b_6}{20b_1^2} + \frac{8b_3b_5}{5b_1^2} 
+ \frac{11b_4^2}{20b_1^2}
- \frac{4b_2b_3b_4}{5b_1^3} + \frac{b_2^2b_5}{10b_1^3} + 
\frac{16b_3^3}{5b_1^3} + \frac{b_2^2b_3^2}{20b_1^4}-\frac{b_2^3b_4}{20b_1^4} 
\quad {\rm for} \ S_{R,m} \ {\rm with} \ m \ge 6 \ , 
\label{k6thooftsol}
\eeq
and
\beqs
k_7 & = & \frac{b_8}{6b_1} 
- \frac{2b_2b_7}{15b_1^2} + \frac{17b_3b_6}{12b_1^2} + \frac{5b_4b_5}{6b_1^2}
+ \frac{b_2^2b_6}{10b_1^3} - \frac{9b_2b_3b_5}{10b_1^3}  
- \frac{49b_2b_4^2}{120b_1^3} + \frac{19b_3^2b_4}{3b_1^3} \cr\cr
& - & \frac{b_2^3b_5}{15b_1^4}-\frac{23b_2b_3^3}{60b_1^4}
+ \frac{9b_2^2b_3b_4}{20b_1^4} 
+ \frac{b_2^4b_4}{30b_1^5} - \frac{b_2^3b_3^2}{30b_1^5} 
\quad {\rm for} \ S_{R,m} \ {\rm with} \ m \ge 7 \ . 
\label{k7thooftsol}
\eeqs
\end{widetext}
%

% =========================================================================

\section{Properties of $S_{R,4}$ Scheme Transformation}
\label{bellprime_sr4}

In this appendix we give some relevant information on the next higher-order
scheme transformation, $S_{R,4}$. The coefficients $b'_\ell$ resulting from the
application of the $S_{R,4}$ scheme transformation are as follows, up to
$\ell=8$ loop order: 
\beq
b'_3 = b'_4 = b'_5 = 0 \ , 
\label{bp345_sr4}
\eeq
\beq
b'_6=b_6-\frac{2b_2b_5}{3b_1}+\frac{8b_3b_4}{b_1}+
\frac{b_2^2b_4}{3b_1^2}-\frac{b_2b_3^2}{3b_1^2} \ ,
\label{bp6_sr4}
\eeq
\beq
b'_7=b_7+\frac{8b_3b_5}{b_1}+\frac{11b_4^2}{4b_1}+
\frac{16b_3^3}{b_1^2}+\frac{2b_2b_3b_4}{b_1^2} \ , 
\label{bp7_sr4}
\eeq
\beqs
b'_8 & = & b_8+\frac{4b_3b_6}{b_1}+\frac{5b_4b_5}{b_1}
-\frac{b_2b_4^2}{4b_1^2}+\frac{2b_3^2b_4}{b_1^2} \cr\cr
& + & \frac{4b_2b_3b_5}{b_1^2}+\frac{12b_2b_3^3}{b_1^3}
-\frac{2b_2^2b_3b_4}{b_1^3} \ .
\label{bp8_sr4}
\eeqs
In general, after the $S_{R,4}$ scheme transformation is applied, the resultant
$n$-loop beta function, $\beta_{\alpha',n\ell}$, has the form of Eq. 
(\ref{betaprime_nloop_srm}) with $m=4$. 

When applied to an asymptotically free gauge theory with $N_f \in I$, so that
there is an IR zero in $\beta_{2\ell}$, the transformation function $f(a')$
evaluated at $a'_{IR,2\ell}=a_{IR,2\ell}=-b_1/b_2$ is 
\beq
f(a'_{IR,2\ell}) = 1 + \frac{b_1 b_3}{b_2^2} - \frac{2b_1^2 \, b_4}{3b_2^3}
+ \frac{5b_1^2 \, b_3^2}{3b_2^4} + \frac{b_1^3 \, b_5}{3b_2^4} \ . 
\label{fap_sr4_air2loop}
\eeq
In order for the $S_{R,4}$ scheme transformation to be acceptable, a necessary
condition is $C_1$, that $f(a') > 0$, in particular, at
$a'=a_{IR,2\ell}'=a_{IR,2\ell}$.  

\end{appendix}

% ========================================================================

% =========================================================================

\newpage

\begin{table}
\caption{\footnotesize{Values of $S_{R,n}$ scheme transformation function
$f(a')$, evaluated at the scheme-independent value of the two-loop IR zero,
$a_{IR,2\ell}'=a_{IR,2\ell}=\alpha_{IR,2\ell}/(4\pi)$, denoted
$f(a')_{IR,S_{R,n}}$.  We list results for $S_{R,n}$ with $n=2$ and $n=3$ in
the SU($N_c$) gauge theory with $2 \le N_c \le 4$ and with $N_f$ fermions
transforming according to the fundamental representation, as functions of $N_c$
and $N_f$, for values of $N_f$ in the respective intervals $I$ where the theory
is asymptotically free and the two-loop beta function $\beta_{2\ell}$ has an
infrared zero.}}
\begin{center}
\begin{tabular}{|c|c|c|c|c|} \hline\hline
$N_c$ & $N_f$ & $\alpha_{IR,2\ell}$ & $f(a')_{IR,S_{R,2}}$ & 
$f(a')_{IR,S_{R,3}}$ \\ 
\hline 
 2  &  7  & 2.83   & $-3.529$   & $-1.898$ \\
 2  &  8  & 1.26   & $-0.5075$  & $-0.154$ \\
 2  &  9  & 0.595  &   0.399    &   0.497  \\
 2  & 10  & 0.231  &   0.795    &   0.813  \\
\hline
 3  & 10  & 2.21   & $-4.454$   & $-3.335$  \\
 3  & 11  & 1.23   & $-1.418$   & $-0.921$  \\
 3  & 12  & 0.754  & $-0.272$   & $-0.027$  \\
 3  & 13  & 0.468  &   0.293    &   0.412   \\
 3  & 14  & 0.278  &   0.616    &   0.667   \\
 3  & 15  & 0.143  &   0.818    &   0.833   \\
 3  & 16  & 0.0416 &   0.9505   &   0.952   \\
\hline 
 4  & 13  & 1.85   & $-5.333$ &  $-4.463$  \\
 4  & 14  & 1.16   & $-2.243$ &  $-1.668$  \\
 4  & 15  & 0.783  & $-0.912$ &  $-0.548$  \\
 4  & 16  & 0.546  & $-0.204$ &  $-0.0207$ \\
 4  & 17  & 0.384  &   0.221  &    0.355   \\
 4  & 18  & 0.266  &   0.498  &    0.573   \\
 4  & 19  & 0.175  &   0.688  &    0.726   \\
 4  & 20  & 0.105  &   0.825  &    0.840   \\
 4  & 21  & 0.0472 &   0.925  &    0.928   \\
\hline\hline
\end{tabular}
\end{center}
\label{fapvalues}
\end{table}

\begin{table}
\caption{\footnotesize{Values of the UV zero in the $\beta$ function of the
U(1) gauge theory with $N_f$ fermions, at $n$-loop ($n\ell$) order, for $n=3, \
4, \ 5$, in the $\overline{\rm MS}$ scheme, denoted $\alpha_{UV,n\ell}$. The
symbol $-$ indicates that there is no zero in $\beta$ for the given order and
value of $N_f$. See text for further details.}}
\begin{center}
\begin{tabular}{|c|c|c|c|} \hline\hline
$N_f$ & $\alpha_{UV,3\ell}$ &$\alpha_{UV,4\ell}$
& $\alpha_{UV,5\ell}$ \\ \hline
 1  & 10.2720   & 3.0400  & $-$      \\
 2  &  6.8700   & 2.4239  & $-$      \\
 3  &  5.3689   & 2.0776  & $-$      \\
 4  &  4.5017   & 1.8463  & $-$      \\
 5  &  3.9279   & 1.67685 & 2.5570   \\
 6  &  3.5156   & 1.5455  & 1.8469   \\
 7  &  3.2027   & 1.4397  & 1.6243   \\
 8  &  2.9555   & 1.3519  & 1.4851   \\
 9  &  2.7545   & 1.2776  & 1.3863   \\
10  &  2.58705  & 1.2135  & 1.3120   \\
\hline\hline
\end{tabular}
\end{center}
\label{alfuvu1}
\end{table} 


\begin{thebibliography}{99}
\bibitem{mf}
%
In a vectorial gauge theory, it is straightforward to include fermion masses,
since these are gauge-invariant.  However, for a given fermion mass $m$, as the
reference scale $\mu$ decreases below $m$, one would integrate these out of the
low-energy effective theory applicable for $\mu < m$, so a massive fermion
would not affect the UV to IR evolution significantly below its mass.  Hence,
to study the IR limit of this evolution, one may restrict to massless
fermions.
\bibitem{chiralgt}
%
In a chiral gauge theory with certain types of fermion content, 
there is also the possibility that the theory confines without any
spontaneous chiral symmetry breaking.
\bibitem{casimir}
%
The Casimir invariants $C_R$ and $T_R$ are defined as $\sum_{a=1}^{o(G)}
\sum_{j=1}^{{\rm dim}(R)}{\cal D}_R(T_a)_{ij} {\cal D}_R(T_a)_{jk} = C_R 
\delta_{ik}$ and $\sum_{i,j=1}^{{\rm dim}(R)}
{\cal D}_R(T_a)_{ij} {\cal D}_R(T_b)_{ji} = T_R \delta_{ab}$, where $R$ is the
representation and $T_a$ are the generators of $G$, so that for SU($N_c$),
$C_A=N_c$ for the adjoint ($A$) and $T_{fund}=1/2$ for the fundamental
representation, etc. $C_f$ denotes $C_R$ for the fermion representation.
\bibitem{nfreal}
%
Here and elsewhere, when expressions are given for $N_f$ that evaluate
to non-integral real values, it is understood that they are formal and are 
interpreted via an analytic continuation of $N_f$ from physical nonnegative 
integer values to real numbers.
\end{thebibliography}
\end{document}